\documentstyle[12pt]{article}
\newcommand{\BE}{\begin{eqnarray}}
\newcommand{\EE}{\end{eqnarray}}
\newcommand{\BF}{\begin{figure}[htbp]}
\newcommand{\EF}{\end{figure}}
\newcommand{\BC}{\begin{center}}
\newcommand{\EC}{\end{center}}
\catcode`\@=11
\def\makepreprititle{\par
  \begingroup
  \def\thefootnote{\fnsymbol{footnote}}
  \def\
@makefnmark{\hbox
  to 0pt{$^{\@thefnmark}$\hss}}
  \if@twocolumn
  \twocolumn[\@makepreprititle]
  \else \newpage
  \global\@topnum\z@
  \@makepreprititle \fi\thispagestyle{empty}\@thanks
  \endgroup
  \setcounter{footnote}{0}
  \let\makepreprititle\relax
  \let\@makepreprititle\relax
  \gdef\@thanks{}\gdef\@author{}\gdef\@title{}
  \gdef\@preprintnumber{}\gdef\@preprintdate{}\gdef\subtitle{}
  \let\thanks\relax}
\def\preprintnumber#1{\gdef\@preprintnumber{#1}}
\def\preprintdate#1{\gdef\@preprintdate{#1}}
\def\subtitle#1{\gdef\@subtitle{#1}}
\def\@makepreprititle{\newpage
{\def\baselinestretch{1}
  \begin{flushright} \@preprintnumber \par
  \@preprintdate \end{flushright} } \par
  \begin{center}
\vskip 1.5em
  {\LARGE \@title \par} \vskip 2.5em
  {\large \lineskip .5em
  \begin{tabular}[t]{c}\@author
  \end{tabular}\par}
  \vskip 1em {\large \@date} \end{center}
  \par
  \vfil}
\date{\sl Department of Physics, Tohoku University\\Sendai, 980 Japan}
\preprintdate{~}
\preprintnumber{~}
\subtitle{}
\def\abstract{\if@twocolumn
\section*{Abstract}
\else \normalsize
\begin{center}
{\bf Abstract\vspace{-.5em}\vspace{0pt}}
\end{center}
\quotation
\addtocounter{page}{-1}
\fi}
\def\endabstract{\if@twocolumn\else\endquotation\fi}
\def\spacing#1{\def\baselinestretch{#1}
\typeout{baselinestretch is modified to \baselinestretch}}
\catcode`\@=12

\spacing{1.3}
\newcommand{\lsim}{\mbox{ \raisebox{-1.0ex}{$\stackrel{\textstyle <}
{\textstyle \sim}$ }}}
\newcommand{\GeV}{\mbox{GeV}}

\renewcommand{\thefootnote}{\fnsymbol{footnote}}
\makeatletter
\long\def\@makefntext#1{\parindent 1em\noindent
 \hbox to 1.8em{\hss${\@thefnmark}$ }#1}
\def\@makefnmark{\hbox{$^{{\@thefnmark}}$}}
\makeatother
\preprintnumber{TU-483}
\preprintdate{May  1995}
\title{Search for Dynamical Symmetry Breaking Physics by Using
Top Quark%
\footnote{Talk given at the 5th JLC Workshop,
Kawatabi, 16-17 February 1995.}
}
\author{T.Asaka$^1$, N.Maekawa$^2$, T.Moroi$^3$\thanks{
Fellow of the Japan Society for the Promotion of Science.},
Y.Shobuda$^1$,
and Y.Sumino$^1$\\
\\
$^1$\sl Department of Physics, Tohoku University\\
\sl Sendai, 980-77 Japan\\
$^2$\sl Department of Physics, Kyoto University\\
\sl Kyoto 606-01, Japan\\
$^3$\sl Theory Group, KEK\\
\sl Tsukuba, Ibaraki 305, Japan}
\date{~}
\begin{document}
\makepreprititle
%
%
%#########################
\begin{abstract}
We report first results from the investigation on probing dynamical
mechanism of the electroweak symmetry breaking using top quark.
We consider the case where the top mass originates from a
fermion-antifermion pair condensate, which necessitates (i)strong
interaction to cause condensation, and (ii)4-fermi interaction to give
top mass.
{}From the observed top mass and the unitarity constraint, we obtain,
for the 4-fermi interaction (ii), lower bounds for its strength and
upper bounds for its intrinsic new mass scale
as we vary the type of
the strong interaction (i).
Bethe-Salpeter and Schwinger-Dyson equations are solved numerically to
study the dynamical symmetry breaking effect semi-quantitatively.
\end{abstract}
%#########################
\clearpage
%
%
%########################
\paragraph{1. Introduction}
%########################
%

{}~

The $SU(2) \times U(1)$ gauge theory for describing the electroweak
interactions has been very successful both theoretically and
experimentally.  However, all the experimental tests have been done only
for its gauge part and  we have little knowledge on
the electroweak symmetry breaking mechanism so far.
In the light of the naturalness problem,
we may suppose that there exists some new physics
related to the electroweak symmetry breaking at the energy scale
{\it O}(100GeV - 1TeV).
Dynamical symmetry breaking is one of the attractive candidates
for the solution to the naturalness problem.
We consider this possibility and intend to study how to
search dynamical symmetry breaking physics at future colliders.
Here we use top quark as a probe for the symmetry breaking sector.
The top quark, being much heavier than other fermions with the mass
of the order of the electroweak symmetry breaking scale
\cite{mt},
is expected to couple most strongly to the
symmetry breaking sector.
In this paper, we assume the gauge symmetry is broken dynamically
by a fermion-antifermion pair condensate and that
the top quark acquires its mass $m_t$
through a 4-fermi interaction
\BE
{\cal L} ~ = ~ \frac{1}{\Lambda^2} \overline{ \Psi}_L \Psi_R
\overline{t}_R t_L + h.c.~,
\label{in4fint}
\EE
where $\Lambda$ represents the new physics scale
and $\Psi$ denotes a new fermion
introduced in the symmetry breaking sector.
Then $m_t$ is estimated to be
\BE
m_t ~ = ~ \frac{\left< \overline{\Psi}_L \Psi_R \right> }{\Lambda^2}.
\label{inmt}
\EE
On the other hand, since the condensate
$\left< \overline{\Psi}_L \Psi_R \right>$ has a non-zero $SU(2)$
charge, it also contributes to the gauge bosons' masses.
Thus we naively expect a following inequality:
\BE
\left<  \overline{ \Psi}_L \Psi_R \right>^{1/3} \leq \Lambda_{EW}
\sim \mbox{{\it O}(250~GeV)},
\label{invev}
\EE
where $\Lambda_{EW}$ represents the electroweak symmetry breaking
scale.
We have inequality for there may exist other
fermion condensations which are $SU(2) \times U(1)$ non-invariant.
{}From the observed value of top mass $m_t \simeq 175 ~ \GeV$
\cite{mt}, this naive argument suggests
the 4-fermi interaction (\ref{in4fint}) is rather strong and
is bounded from below
$1/\Lambda^2 \geq m_t/\Lambda^3_{EW}$
due to eqs.(\ref{inmt}) and (\ref{invev}).
It means that the new physics scale
$\Lambda$ is not too far from $\Lambda_{EW}$.
This is the reason why we use the top quark (or the 4-fermi interaction
(\ref{in4fint})) to probe the symmetry breaking
mechanism.
{}From the non-standard effect to
e.g.\ $Z\overline{t}t$ vertex as shown in fig.\ref{ztt},
we would be able to study dynamical symmetry breaking physics at
future colliders.
%
%
%\BF
%\BC
%\epsfile{file=ztt.eps,height=4cm}
%\caption[]{Diagram representing the non-standard correction
%to the $Z \overline{t}t$ vertex
%from the dynamical symmetry breaking sector.}
%\label{ztt}
%\EC
%\EF
%
%

%
In this note, we report, as our first results, the numerical
estimation of a 4-fermi interaction which gives rise to the top mass
dynamically.
In this analysis, we assume a strong interaction which causes the
fermion and anti-fermion pair condensate,
and a 4-fermi interaction to give the top mass.
For the strong interaction, we examine the
non-abelian gauge interaction cases such as $SU(2)$ or $SU(3)$
and the fixed coupling constant case.
{}From the experimental value of $m_t$ and the unitarity constraint,
we obtain lower bounds for the strength of the 4-fermi interaction
and upper bounds for its intrinsic new mass scale.

{}~

%
%
%
%
%###################################################################
\paragraph{2. Numerical Estimation of the 4-Fermi Interaction}
%###################################################################
%

{}~

First, we explain the electorweak symmetry breaking sector
considered here.
We introduce fermions to the breaking sector following the
one-doublet technicolor (TC) model\cite{tc1,tc2},
\BE
Q_L  = \left( \begin{array}{c} U \\ D \end{array} \right)_L,
{}~~~~~~~~
U_R ,
{}~~~~~~~~
D_R ,
\EE
with the weak hypercharge,
\BE
Y(Q_L) ~ = ~ 0 ,
{}~~~~
Y(U_R) ~ = ~ \frac{1}{2},
{}~~~~
Y(D_R) ~ = ~ -\frac{1}{2} ~,
\EE
where these quantum numbers are chosen to cancel anomaly.
$N_d$ sets of above new fermions are introduced in the breaking
sector.
In order to characterize the strong interaction between the above
fermion and antifermion, we introduce kernel $K$.
For example, in the TC models where the $SU(N_{TC})$ gauge
interaction cause the condensation,
we can approximately write down the explicit form of the kernel $K$.
In the improved ladder approximation in Landau gauge
\cite{improved-ladder},
it is written in the momentum space as
\cite{fpi}
(momentum configuration is defined in fig.\ref{kernel-fig})
\BE
K(p,q) ~ = ~ C_2 g^2(p,k) ~
\left(\gamma^{\mu} \otimes \gamma^{\nu}\right)
\frac{1}{(p-k)^2}
\left[ g_{\mu \nu} - \frac{ (p-k)_{\mu} (p-k)_{\nu} }{(p-k)^2}
\right]
{}~,
\label{kernel}
\EE
where $C_2$ is the second Casimir invariant of the $SU(N_{TC})$ fundamental
representation and $g(p,k)$ is the running coupling constant.
%
%
%\BF
%\BC
%\epsfile{file=kernel.eps,height=2.5cm}
%\caption[]{Feynman diagram for the kernel $K(p,k)$.
%The helix denotes the techni-gluon propagator.}
%\label{kernel-fig}
%\EC
%\EF
%
%
We can deal with various types of strong interaction
by changing the form of the kernel $K$.
Here we consider two possibilities.
One is the non-abelian gauge interaction ($SU(2)$ and $SU(3)$) case
and the other is the fixed coupling case.
Once the strong interaction to cause a condensate is chosen and
the form of the kernel is fixed,
we can write down the SD eq.\ and BS eq.
%which are the self-consistent equations for the full 2-point and 4-point
%Green functions of $\Psi$, respectively.
These equations in the ladder approximation
are shown diagrammatically in fig.\ref{sdbs-fig}.
%
%\BF
%\BC
%\epsfile{file=fsdbs.eps,height=4cm}
%\caption[]{The graphical representations of the
%SD eq.\  and the BS eq.\  in the ladder approximation.
%The line with a blob denotes the full propagetor of $\Psi$.}
%\label{sdbs-fig}
%\EC
%\EF
%
%
By solving these equations numerically,
we obtain the full propagator of $U$ and $D$,
and the decay constant $F_{\pi}$, and so on \cite{fpi,meson}.
Note that this approach have been applied to for the QCD case
and the results of numerical calculation are in good agreement
with the experimental values \cite{fpi,meson}
\footnote{
%By using the numerical solutions of these equations,
%we can calculate the dynamical symmetry breaking effects
%semi-quantitatively.
%To study the dynamical symmetry breaking effect,
%we solve numerically the Schwinger-Dyson (SD) and Bethe-Salpeter (BS)
%equations.
Taking $f_{\pi}$ as the only input parameter,
$\Lambda_{QCD}$,$\left< \overline{\Psi}\Psi \right>$,
and various meson masses and decay constants
have been calculated in refs.\cite{fpi,meson}.
So we expect to calculate the top mass,
the oblique corrections, the $Z\overline{t}t$ vertex,
etc., semi-quantitatively.
}.
%
%
%
%============================
\subparagraph{Top Quark Mass}
%============================
%
%
%
Here we calculate the top mass following the approach
discussed above.
We introduce a 4-fermi interaction which induces the top mass,
\BE
{\cal{L}} = \frac{G}{M^2}
( \overline{Q_L} U_R ) ( \overline{t_R} q_L ) + h.c.~,
\label{4fint}
\EE
where $q_L$ denotes the ordinary quark weak doublet
and $G$ the dimensionless coupling.
Because the 4-fermi interaction (\ref{4fint})
cannot be a fundamental interaction,
there should be a mass scale above which this interaction will
resolve.
We call this new physics scale $M$.
Therefore the 4-fermi interaction (\ref{4fint}) should be
regarded as an effective interaction which is appropriate
below energy scale $M$
\footnote{
For the models such as the extended TC model\cite{etc},
various 4-fermi interaction other than (\ref{4fint})
are induced below energy scale $M$.
We do not consider such a possibility.
In this article we only introduce the 4-fermi interaction
(\ref{4fint}) by hand.
}.
Now the top quark acquires its mass through the 4-fermi interaction
(\ref{4fint}) as shown in fig.\ref{figtmass}.
%
%
%\BF
%\BC
%\epsfile{file=ftmass.eps,width=4cm}
%\caption[]{The diagram for the top quark mass.}
%\label{figtmass}
%\EC
%\EF
%
%
Using the full propagator of $U$ which is a solution to the
 SD equation
\footnote{
In fact, the SD eq. shown in fig.\ref{sdbs-fig} is incomplete.
Because of the 4-fermi interaction (\ref{4fint}),
we should consider the mixing of SD eqs. of top quark and
$U$, but here we neglect this effect.
},
the top mass can be calculated as
\BE
m_t = \frac{G}{M^2}
\left< \overline{U}U \right>_M~,
\label{topmass}
\EE
where
\BE
\left< \overline{U}U \right>_M &=& \frac{1}{2}
		 \int^M \frac{d^4 p}{( 2 \pi)^4}
		\mbox{tr}
		\left( \frac{i}{ \not\!p - \Sigma(p) } \right)
\nonumber
\\
	&=& \frac{1}{2} \frac{ N_{TC} N_d }{ 4 \pi^2}
		\int^{M^2} dx
		\frac{ x \Sigma(x) }{ x + \Sigma(x)^2 }
\label{vev}
\EE
with $x=p_E^2 =-p^2$.
Because above the scale $M$ the 4-fermi interaction will resolve,
we set the upper bound of the integral as $M$
\footnote{
In fact the new physics scale $M$ is defined as the upper bound
of this integral in this article.}
{}.
Using the decay constant $F_{\pi}$
\footnote{
We define the decay constant $F_{\pi}$ as
$
\left< 0 \right|
\overline{\Psi}T^a \gamma^{\mu} \gamma^5 \Psi
\left| \pi^a(p) \right> =  - i p^{\mu} F_{\pi},
$
where $\Psi = (U,D)$.
}
which is obtained by
solving the BS eq.,
we can set the mass scale of the theory.
{}From the gauge bosons masses, $F_{\pi}$ is normalized as
\BE
M_W ~ \geq ~ \frac{gF_{\pi}}{2},
\label{mass-scale}
\EE
where $g$ is the $SU(2)_L$ gauge coupling constant.
Here we have inequality rather than equality,
because there may be fermion condensates other than
$\left<\overline{U}U \right>$ and
$\left<\overline{D}D \right>$
which do not contribute to the top mass while giving masses
to the gauge bosons.
By substituting the observed top mass $m_t \simeq 175 ~ \GeV$
\cite{mt} in eq.(\ref{topmass}),
we obtain the coupling $G$ for a given $M$.
We consider the case where the strong interaction
is given by the $SU(2)$ and $SU(3)$
gauge interaction, and
the constraints on $G$ are given as the function of $M$
in figs.\ref{su2} and \ref{su3}, respectively.
In each figures, the upper region of the curved lines are allowed
because of the inequality in eq.(\ref{mass-scale}).
The coupling $G$ is found to be bounded from below.
The behavior of the lower bound of $G$ can be understood
by considering the momentum dependence of the mass function
$\Sigma(x)$.
For $x\ll \Lambda_{TC}^2$
($\Lambda_{TC}$ is defined as the scale where the
leading-logarithmic running coupling constant diverges),
$\Sigma(x)$ is almost flat. This gives the asymptotic
behavior of the lower bound as a function of $M$,
\BE
&G& ~\sim ~\frac{1}{M^2}.
\EE
The lower bound of $G$ increases rapidly as $M\rightarrow0$,
because the upper bound of the integral in eq.(\ref{vev})
becomes smaller.
However, we neglect this region of $M$ ( $M \ll \Lambda_{TC}$ )
because our effective treatment of
the 4-fermi interaction (\ref{4fint}) becomes invalid.
On the other hand, for $x \gg \Lambda_{TC}^2$,
the mass function decreases rapidly as
\BE
\Sigma(x) ~ \sim ~ \frac{1}{x} \left( \ln x \right)^{
\frac{1}{4 B} - 1 }~,
\EE
where $B = \beta_0/12 C_2$ with $\beta_0$ being the lowest order
coefficient of the $\beta$ function.
{}From this momentum dependence, the lower bound behaves as
\BE
&G& ~\sim ~\frac{M^2}{ (\ln M^2)^{\frac{1}{4B}}}.
\label{lower}
\EE
Therefore the lower bound for coupling $G$ is found to be
approximately proportional to $M^2$ and the lower
bound for the strength of the
4-fermi interaction (\ref{4fint}), typically $\sim G/M^2$,
is obtained.
This result agrees with the naive discussion given
in the introduction.
We also consider the fixed coupling constant case
( $B$ = 0 )
in order to apply to the QED like theory or
walking technicolor theory\cite{wtc}.
In fig.\ref{wtc} we show the constraints for the coupling $G$.
In this case, as $x \rightarrow \infty$,
the decreasing behavior of $\Sigma(x)$ is different
from the former case, and is
\BE
\Sigma(x) ~\sim~ \frac{1}{\sqrt{x}}.
\EE
{}From this dependence the lower bound of the coupling $G$
increases linearly as the function $M$
and the lower bound of the strength is weaker than the one in
former non-abelian gauge interaction cases.
%
%
%
%
%============================
\subparagraph{Unitarity Constraint}
%============================
%
%
Next, we consider the unitarity constraint and put upper
bounds for $G$.
First of all, we explain the unitarity bounds which are used here.
Let us consider a two-body to two-body scattering
( 1 + 2 $\rightarrow$ 3 + 4 ).
We take the helicity of each particle as $\lambda_1$,$\lambda_2$,
$\lambda_3$, and $\lambda_4$.
In the center of mass (C.M.) frame the partial-wave expansion for the
helicity amplitude is,
\BE
M_{fi} ~ = ~
\frac{8\pi}{\sqrt{{\beta_i}{\beta_f}}}
\sum^{\infty}_{0} ( 2 J + 1 )
T^J_{\lambda_1,\lambda_2;\lambda_3,\lambda_4}(\sqrt{s})
{}~d^J_{\lambda_i,\lambda_f}(\theta)
{}~\mbox{e}^{i ( \lambda_i - \lambda_f ) \phi }
,
\EE
where $\lambda_i = \lambda_1 - \lambda_2$,
$\lambda_f = \lambda_3 -\lambda_4$, $\beta_i = 2 |{\bf{p_i}}|/\sqrt{s}$,
$\beta_f = 2 |{\bf{p_f}}|/\sqrt{s}$ and
$d^J_{\lambda_i,\lambda_f}(\theta)$ is the Wigner's $d$-function.
And ${\bf{p_i}}({\bf{p_f}})$ is the common C.M. momentum for the initial
(final) state and $\sqrt{s}$ the C.M. energy.
The angle $(\theta,\phi)$ is taken as the direction angle
of the final particle 3 measured from the direction of the initial
particle 1.
{}From the unitarity condition for the $S$-matrix
$S^{\dagger}S = S S^{\dagger} = 1$,
the unitarity bounds for the coefficient of the
partial-wave expansion
$T^J_{\lambda_1,\lambda_2;\lambda_3,\lambda_4}(\sqrt{s})$
are obtained as
\BE
&&|\mbox{Re} T^J| ~ \leq ~ 1 ~~~~~( \mbox{for an elastic channel} )
\\
&&|T^J| ~ \leq ~ 1 ~~~~~~~( \mbox{for an inelastic channel} )
\label{ub2}
\EE
Using these unitarity bounds, we put the upper bound of $G$.
We consider the two-body to two-body scattering of fermions
through the 4-fermi interaction (\ref{4fint}) at the energy scale
$E \gg \Lambda_{TC}$ where the effect of confinement
can be neglected.
Matrix elements of these processes grow as $\sim E^2$ as
the scattering energy increases, and it would break the above
unitarity bounds.
But above the energy scale $M$, the 4-fermi interaction
should resolve to retain unitarity.
Thus the upper bound for $G$ is obtained by assuming that
the unitarity condition would not break down below the scale $M$.
We found that the most stringent bound comes from
the scattering process
$t\overline{t} \rightarrow U\overline{U}$.
{}From the 4-fermi interaction (\ref{4fint}),
this process occurs only in a scalar channel ($J$=0).
%The helicity amplitude for this process
%
%
%\BE
%M^0_{\frac{1}{2},\frac{1}{2};\frac{1}{2},\frac{1}{2}} ~=~
%M^0_{-\frac{1}{2},-\frac{1}{2};-\frac{1}{2},-\frac{1}{2}} ~=~
%\frac{G}{M^2} ~s
%\EE
%
%
The coefficient of the partial-wave expansion
in the massless limit $m_t = m_U = 0$ is given by
\BE
T^0(\sqrt{s}) ~ =~ \frac{1}{8\pi} ~\frac{G ~s}{M^2}.
\EE
By the above assumption, the upper bound for the coupling $G$
comes from
\BE
|T^0(\sqrt{s} =M)| ~\leq ~ 1,
\EE
and this leads,
\BE
G ~\le ~ 8 \pi.
\EE
This bound for the coupling $G$ is shown
in figs.\ \ref{su2}, \ref{su3}, and \ref{wtc}.
%
%
%
%============================
%\subparagraph{Results}
%============================
%
%
%
%
%We have discussed about the 4-fermi interaction which is needed to
%explain the top quark mass.
%Here we represent the results obtained by above discussion.
%First, from the observed top mass we obtain the lower bounds
%for the strength of the 4-fermi interaction in eq.(\ref{4fint})

Combining the previous estimation of the top mass
and this unitarity constraint,
we obtain the allowed region in the $G$--$M$ plane for each cases.
{}From these figures the typical value of the coupling $G/4\pi$ is
found to be {\it O}(1) in all cases and
to be a rather strong coupling constant.
Moreover, we can find the upper bounds for the new physics scale
$M$ in the 4-fermi interaction (\ref{4fint}).
For the $SU(2)$ and $SU(3)$ gauge interaction case,
the upper bounds are around 5 TeV.
For the fixed coupling constant case,
the bound is much weaker than the former case and
is around 15 TeV.

\clearpage
%
%###########################################
\paragraph{3. Conclusions and Discussion}
%###########################################

{}~

We have considered the case where the top quark acquires mass
from a fermion-antifermion condensate.
In this case, we assume a strong interaction to cause condensation
and a 4-fermi interaction to give top mass.
We consider two possibilities for the strong interaction.
One is the non-Abelian gauge interaction ( $SU(2)$ and $SU(3)$ ) case.
The other is the fixed coupling constant case.
{}From the observed top quark mass and the unitarity constraint,
we obtain the following results for the 4-fermi interaction in
each case.
\begin{itemize}
	\item{Lower bound for its strength is obtained.}
	\item{Coupling $G/4\pi$ is typically of {\it O}(1).}
	\item{Upper bound for the new physics scale $M$ is obtained:}
\begin{description}
	\item{$\cdot$ $M$ $\lsim$ 5 TeV for the non-Abelian gauge interaction
		case.}
	\item{$\cdot$ $M$ $\lsim$ 15 TeV for the fixed coupling case.}
	\end{description}
\end{itemize}
The above results will be the bases in studying the dynamical
symmetry breaking mechanism by using top quark.
{}From these results, we intend to calculate the non-standard
corrections to the $\gamma \gamma \overline{t}t$ vertex,
the $Z\overline{t}t$ vertex, the $W\overline{t}b$ vertex and so on.
%Solving SD eq.\ and BS eq.\ numerically,
%we can analyze these non-perturbative effects semi-quantitatively.
Our goal will be to discuss the detectability of these corrections
at future $e^+e^-$ and hadron colliders,
and probe the dynamical symmetry breaking mechanism
by using top quark. This will be given elsewhere\cite{asaka}.
\clearpage
%###############################################################

\clearpage
%###############################################################
\section*{Figure Caption}

{}~

%
%###########################################
\newcounter{asaka}
\begin{list}%
{Figure \arabic{asaka}:}{\usecounter{asaka}
\setlength{\labelsep}{1cm}
\addtolength{\labelwidth}{2cm}
\addtolength{\leftmargin}{2cm}
\addtolength{\itemsep}{0.5cm}}
\item{Diagram representing the non-standard correction
      to the $Z \overline{t}t$ vertex
      from the dynamical symmetry breaking sector.}
\label{ztt}
%%###########################################
\item{Feynman diagram for the kernel $K(p,k)$.
      The helix denotes the techni-gluon propagator.}
\label{kernel-fig}
%###########################################
\item{The graphical representations of the
      SD eq.\  and the BS eq.\  in the ladder approximation.
The line with a blob denotes the full propagetor of $\Psi$.}
\label{sdbs-fig}
%###########################################
\item{The diagram for the top quark mass.}
\label{figtmass}
%###########################################
\item{Allowed region in the $G$-$M$ plane for $m_t$ = 175 GeV.
The $SU(2)$ gauge interaction is considered as the strong
attractive force
to cause the condensation.
The lower bound for $G$ from the top quark mass is shown with curved lines
in the case where the number of the set of $U$ and $D$,
$N_d$ is taken as $N_d$ = 1, 2, and 3.
In this case, $\Lambda_{TC}$ = 1.7, 1.1, and 0.7 TeV for
$N_d$ = 1,2, and 3, respectively.
The upper bound for $G$ from the unitarity constraint is also shown.
{}From this figure, we can obtain the upper bounds for the scale $M$.}
\label{su2}
%###########################################
%
%
\item{Allowed region in the $G$-$M$ plane for $m_t$ = 175 GeV.
The $SU(3)$ gauge interaction is considered as the strong
attractive force
to cause the condensation.
The lower bound for $G$ from the top quark mass is shown with curved lines
in the case where the number of the set of $U$ and $D$,
 $N_d$ is taken as $N_d$ = 1, 2, and 3.
In this case, $\Lambda_{TC}$ = 1.3, 0.9, and 0.7 TeV for
$N_d$ = 1, 2, and 3, respectively.
The upper bound for $G$ from the unitarity constraint is also shown.
{}From this figure, we can obtain the upper bounds for the scale $M$.}
\label{su3}
%###########################################
%
%
\item{Allowed region in the $G$-$M$ plane for $m_t$ = 175 GeV.
The interaction of a fixed coupling constant considered as
the strong attractive force to cause the condensation.
The lower bound for $G$ from the top quark mass is shown with curved
lines
in the case where the number of the set of $U$ and $D$,
 $N_d$ is taken as $N_d$ = 3, 6, and 9.
The upper bound for $G$ from the unitarity constraint is also shown.
{}From this figure, we can obtain the upper bounds for the scale $M$.}
\label{wtc}
\end{list}
%###########################################
%
\end{document}